\documentclass[aps,twocolumn,showpacs,pra,a4paper,floatfix,nofootinbib]{revtex4-1}
\usepackage{amsmath}
\usepackage{amssymb}
\usepackage{amsfonts}
\usepackage{fancybox}
\usepackage{eepic}
\usepackage{times}
\usepackage{latexsym}
\usepackage{pifont}
\usepackage{graphicx}
\usepackage{amstext}

\def\acetylene{C$_2$H$_2$}
\def\jpbo{J. Phys. B: At. Mol. Opt. Phys.}   

\def\pra{Phys. Rev. A}

\def\prl{Phys. Rev. Lett.}

\def\jcp{J. Chem. Phys.}        
\newcommand{\bv}[1]{\mbox{\boldmath $#1$}}
\newcommand{\intensity}[2]{$#1\times10^{#2}\mbox{ W/cm}^{2}$}
\def\md{\mathrm{d}}

\begin{document}

\title{High-order harmonic generation from highly-excited states in acetylene}
\author{Peter Mulholland and Daniel Dundas}
\affiliation{Atomistic Simulation Centre, School of Mathematics and Physics, Queen's University Belfast, 
University Road, Belfast BT7 1NN. N. Ireland.}
\date{\today}

\begin{abstract}
High-order harmonic generation (HHG) from aligned acetylene molecules interacting with mid infra-red (IR), 
linearly polarized laser pulses is studied theoretically using a mixed quantum-classical approach in which 
the electrons are described using time-dependent density functional theory while the ions are treated classically.
We find that for molecules aligned perpendicular to the laser polarization axis, HHG arises from the highest-occupied
molecular orbital (HOMO) while for molecules aligned along the laser polarization axis, HHG is dominated by the 
HOMO-1. In the parallel orientation we observe a double plateau with an inner plateau that is produced by ionization 
from and recombination back to an autoionizing state. Two pieces of evidence support this idea. Firstly, by choosing 
a suitably tuned vacuum ultraviolet pump pulse that directly excites the autoionizing state we observe a dramatic 
enhancement of all harmonics in the inner plateau. Secondly, in certain circumstances, the position of the inner 
plateau cut-off does not agree with the classical three-step model. We show that this discrepancy can be understood 
in terms of a minimum in the dipole recombination matrix element from the continuum to the autoionizing state.
As far as we are aware, this represents the first observation of harmonic enhancement over a wide range of 
frequencies arising from autoionizing states in molecules.
\end{abstract}

\pacs{42.65.Ky,33.80.Rv,31.15.ee,31.15.xf}
\maketitle

\section{Introduction}
High-order harmonic generation (HHG) is a highly non-linear process in which an atom or molecule absorbs 
energy from an intense laser pulse before emitting short attosecond bursts of radiation with a frequency 
that can be many multiples of the incident laser frequency. Understanding and controlling HHG is crucial 
since it provides a versatile tool for a range of different applications including the use of high harmonic 
spectroscopy to probe chemical reactions~\cite{marangos:2016}, using it as a tool for imaging individual 
molecular orbitals~\cite{itatani:2004} and for producing attosecond pulse trains~\cite{krauz:2009}. HHG is 
usually described using the classical three-step model in which an electron is ionized by the laser pulse, 
is subsequently driven far from the parent ion by the field, before finally recombining with the parent with 
the resultant emission of radiation~\cite{corkum:1993,kulander:1993}.

Generally, it is assumed that the electron ionizes from and returns to the ground state. However, many studies 
now propose different schemes involving excitation to intermediate states that can greatly influence the 
generation of harmonics. For example, early studies of HHG from a coherent superposition of states showed 
that multiple plateaus can be produced with cut-offs associated with the ionization potential of the states 
involved~\cite{watson:1996}. Four-step models of HHG have also been developed to describe how autoionizing 
states can influence the harmonic response~\cite{strelkov:2010}. In addition, studies of HHG in asymmetric 
diatomic molecules have observed double plateau structures which have been ascribed to resonant excitation 
induced by laser-induced electron transfer~\cite{bian:2010}. In that case, HHG is associated with a channel 
in which electrons are ionized from an excited state and recombine to the ground state. Recently, this channel 
has been observed experimentally in studies of HHG in argon atoms~\cite{beaulieu:2016}. In many cases the 
enhancement is only observed over a narrow band of harmonics~\cite{ganeev:2006,rothhardt:2014}. However, by 
exciting atomic targets with an IR pulse in conjunction with a high-order harmonic pulse, a dramatic enhancement 
of the harmonic spectra across a wide frequency range was observed~\cite{ishikawa:2003,takahashi:2007}. 

In this work we show that highly-excited states play an important role in HHG from aligned acetylene 
molecules interacting with mid infra-red (IR) linearly polarized laser pulses ($\lambda =$ 1450 nm). Using 
such long laser wavelengths allows the production of an extended plateau using laser intensities well below 
the saturation limit~\cite{marangos:2016}. Indeed, several studies of HHG in acetylene have already been 
carried out~\cite{vozzi:2010,torres:2010,vozzi:2012,negro:2014}. These have mainly concentrated on studying 
the role of structural interference minima~\cite{lein:2002a,lein:2002b}. Acetylene (\acetylene) is a small, 
linear polyatomic molecule with the ground state configuration 
$(1\sigma_g)^2(1\sigma_u)^2(2\sigma_g)^2(2\sigma_u)^2(3\sigma_g)^2(1\pi_u)^4$. In addition, the next lowest 
unoccupied orbitals  are $(1\pi_g)^0(3\sigma_u)^0(4\sigma_g)^0$. It is widely known that resonance phenomena 
greatly influence photoabsorption and photoionization cross sections in 
acetylene~\cite{collin:1967,langhoff:1981,zamith:2003,machado:1982,han:1989,mitsuke:1994,wells:1999,%
yasuike:2000,fronzoni:2004}. While many studies have considered excitations from the highest occupied 
molecular orbital (HOMO) to the lowest unoccupied molecular orbital (LUMO), there is much interest in the role 
of highly-excited states in photoabsorption and photoionization at vacuum ultraviolet (VUV) 
wavelengths~\cite{machado:1982,han:1989,mitsuke:1994,wells:1999,yasuike:2000,fronzoni:2004}. Therefore, we 
can envisage that these highly-excited states can give rise to several possible mechanisms for HHG.

The paper is arranged as follows. In Sec.~\ref{sec:method} we describe our approach for describing HHG in 
acetylene, namely time-dependent density functional theory (TDDFT). In particular, we detail the various 
numerical parameters used in the simulations reported here. Sec.~\ref{sec:results} then presents our results. 
Firstly, we show how our calculations give an accurate description of the static properties of acetylene, in 
terms of ionic configurations, ionization potentials, excited states, etc. Then we present calculations of 
HHG in acetylene using a range of short-duration laser pulses with wavelengths ranging from VUV to mid-IR 
wavelengths. These simulations show that populating particular highly-excited states using a VUV pulse greatly 
enhances the harmonic yield produced through interaction with a mid-IR pulse. In addition, we present the 
populations in a range of states in order to provide information on those states that are excited during 
interaction with the laser pulses. Finally, in Sec.~\ref{sec:conclusions} some conclusions are drawn.

Unless otherwise stated, atomic units are used throughout.

\section{Method}
\label{sec:method}
Our calculations are carried out using TDDFT~\cite{runge:1984}, as implemented in our code 
EDAMAME~\cite{dundas:2012,wardlow:2016}. EDAMAME is a highly-parallelized implementation of the nonadiabatic 
quantum molecular dynamics method, in which the electronic dynamics are calculated quantum mechanically using 
TDDFT on a real-space grid, while the ionic motion is treated classically. While TDDFT has well-known problems 
in describing autoionizing resonances arising from double excitations, it does capture those arising from 
single excitations~\cite{krueger:2009}. For the resonance phenomena studied in this work, previous TDDFT studies 
of photoionization of acetylene have shown good accuracy~\cite{fronzoni:2004}.

The electronic dynamics are modelled by solving the time-dependent Kohn-Sham equations (TDKS)
\begin{equation}
i\frac{\partial }{\partial t}\psi_{j} (\bv{r},t) = H_{\mbox{\scriptsize ks}}\psi_{j} (\bv{r},t) 
\hspace*{1cm} j = 1,\dots, N,
\label{eq:tdks}
\end{equation}
with the Kohn-Sham Hamiltonian $H_{\mbox{\scriptsize ks}}$ defined as
\begin{equation}\label{eq:ks_hamiltonian}
 H_{\mbox{\scriptsize ks}} = \Biggr [ 
-\frac{1}{2} \nabla^2 + V_{\mbox{\scriptsize H}}(\bv{r}, t)+
V_{\mbox{\scriptsize ext}}(\bv{r}, \bv{R}, t)+ V_{\mbox{\scriptsize xc}}(\bv{r},  t) \Biggr],
\end{equation}
where $V_{\mbox{\scriptsize H}}(\bv{r}, t)$ is the Hartree potential, 
$V_{\mbox{\scriptsize ext}}(\bv{r}, \bv{R}, t)$ is the external potential, 
$V_{\mbox{\scriptsize xc}}(\bv{r},  t)$ is the exchange-correlation  potential, and $\psi_{j} (\bv{r},t)$ 
are the Kohn-Sham orbitals. Neglecting spin effects, the time-dependent electron density is then calculated as 
\begin{equation}
n(\bv{r},t) =  2\sum_{j=1}^{N} \left | \psi_{j } (\bv{r},t)  \right |^2,
\end{equation}
where each Kohn-Sham orbital has an initial occupation of two, and $N = N_e / 2$, where $N_e$ is the 
number of active electrons (for clarity here we have neglected spin degrees of freedom).

The exchange-correlation potential, $V_{\mbox{\scriptsize xc}}(\bv{r},  t)$, in Eq.\,(\ref{eq:ks_hamiltonian}), 
which accounts for all electron-electron interactions in the TDKS equations, is approximated using the local 
density approximation incorporating the Perdew-Wang parameterization of the correlation 
functional~\cite{Perdew:1992}. This functional contains self-interaction errors which means that its long 
range behaviour is incorrect. One major consequence is that electrons are too loosely bound and excited states 
are not accurately described~\cite{casida:1998}. Therefore, we supplement this functional with the average 
density self-interaction correction (ADSIC)~\cite{legrand:2002} which reinstates the correct long-range 
behaviour in an orbital-independent fashion.

The Kohn-Sham orbitals in EDAMAME are discretised on 3D finite difference grids in Cartesian coordinates. 
The grid spacing was taken to be 0.4 a$_0$ in each direction and the grid extents were:  
$|x| = |y| \leq 90.8$\,a$_0$ and  $|z| \leq 146.8$\,a$_0$. Absorbing boundaries are implemented near the edges 
of the grid using a mask function technique~\cite{dundas:2012} in order to prevent wavepacket reflections from 
the edges of the grid.

The external potential $V_{\mbox{\scriptsize ext}}(\bv{r}, \bv{R}, t)$ accounts for both the interaction of 
the laser field with the electrons, and the electron-ion interactions. The laser-interaction term is described 
within the dipole approximation in the length gauge. The electric field vector, $\bv{E}(t)$, is defined through 
its relationship to the vector potential, $\bv{A}(t)$, namely 
\begin{equation}
\bv{E}(t) = -\frac{\partial}{\partial t}\bv{A}(t). 
\end{equation}
All calculations presented in this paper consider acetylene interacting with linearly polarized pulses 
described by $\sin^2$ pulse envelopes~\cite{dundas:2012}
\begin{equation}
 f(t) = 
 \begin{cases}
  \sin^2\left(\frac{\displaystyle \pi t}{\displaystyle T}\right) &0 \leq t \leq T \\
  0 &\textrm{otherwise},
 \end{cases}
\end{equation}
where $T$ is the pulse duration. In this case the vector potential is defined as
\begin{equation}
\label{eq:vector_potential}
\bv{A}(t) = A_0 f(t) \cos(\omega_L t)\hat{{\bv{e}}},
\end{equation}
%
from which $\bv{E}(t)$ can be written as 
\begin{equation}
\bv{E}(t)= \left( E_0 f(t) \sin(\omega_L t) - 
\frac{E_0}{\omega_L}\frac{\partial f}{\partial t} \cos (\omega_L t)\right) \bv{\hat{e}}.
\end{equation}
Here, $A_0$ and $E_0$ are the peak values of the vector potential and electric field vector respectively, 
$\omega_L$ is the laser frequency and $\hat{\bv{e}}$ is the unit vector in the polarization direction of 
the laser field.

For the pulses considered, we assume the innermost electrons do not contribute to the dynamical response 
and therefore only consider the response of the ten outermost electrons. The electron-ion interactions are 
described with Troullier-Martins pseudopotentials~\cite{troullier:1991} in the Kleinman-Bylander 
form~\cite{Kleinman:1982}. All pseudopotentials were generated using APE (the Atomic Pseudopotentials 
Engine)~\cite{Oliveira:2008}. 

The ionic motion is treated classically using Newton's equations of motion. For ion $k$ we have
\begin{align}
M_k \ddot{\bv{R}}_k = &-\int n (\bv{r},t) \frac{\partial H_{\mbox{\scriptsize ks}}}{\partial \bv{R}_k} 
\md\bv{r} \nonumber \\
& -\frac{\partial}{\partial \bv{R}_k} \Bigr(V_{nn} (\bv{R}) + Z_k \bv{R}_k\cdot \bv{E}(t)\Bigr),
\label{eq:neom}
\end{align}
where $V_{nn} (\bv{R})$ is the Coulomb repulsion between the ions and $Z_k \bv{R}_k\cdot \bv{E}(t)$ 
denotes the interaction between ion $k$ and the laser field. 

The TDKS equations, Eq.~(\ref{eq:tdks}), are propagated in time using an 18th-order Arnoldi propagator, 
while the ionic equations of motion (\ref{eq:neom}) are propagated in time using the velocity-Verlet 
method. The timestep in both cases was of 0.2\,a.u..

\begin{figure}[t]
\centerline{\includegraphics[width=7cm,viewport=57 21 559 545]{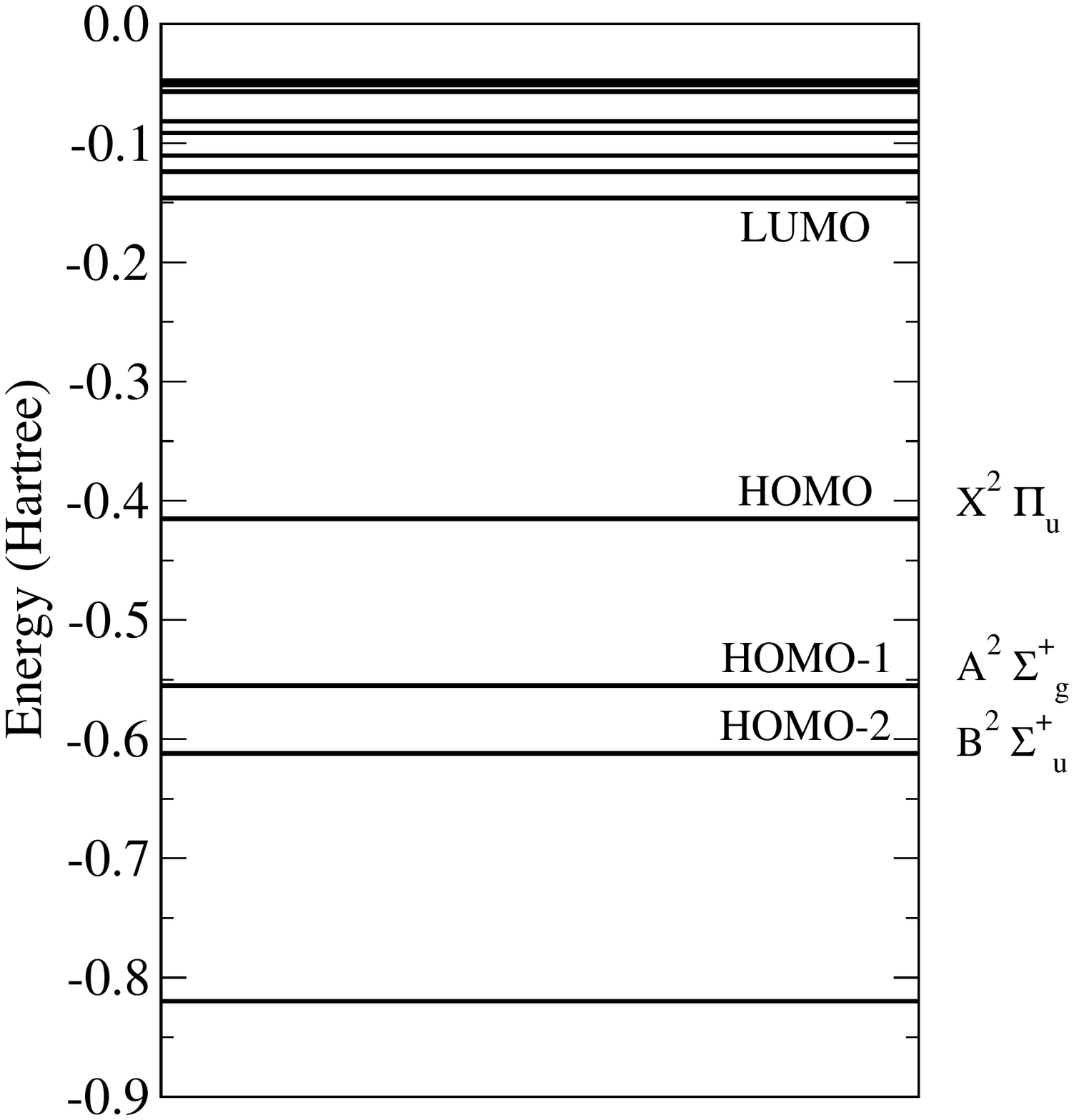}}
\caption{Occupied and unoccupied Kohn-Sham orbital energies of acetylene, obtained from a geometry-relaxed 
calculation of the ground state using the ADSIC exchange-correlation functional. From Koopman's theorem, 
the magnitude of the HOMO, HOMO-1 and HOMO-2 energies approximate ionization potentials to the ground 
state ($X^2 \Pi_u$, 0.4191~Ha ), first excited state ($A ^2\Sigma_g^+$, 0.6140~Ha) and second excited state 
($B ^2\Sigma^+_u$, 0.6912~Ha) of the acetylene cation respectively.}
\label{fig:ks_energies}
\end{figure}

\section{Results}
\label{sec:results}
In this section we present results for the interaction of acetylene with intense laser pulses. In
particular we consider the harmonic response of the molecule while at the same time considering the 
excitation and ionization dynamics. Firstly, we describe how these quantities are calculated
using our approach.

We calculate HHG spectra by taking the Fourier transform of the dipole acceleration, 
along the direction $\hat{\bv{e}}_k$~\cite{burnett:1992}
\begin{equation}\label{eq:hhg_all}
S_k(\omega) = \left | \int e^{ i \omega t } \,\hat{\bv{e}}_k \cdot \ddot{\bv{d}}(t)\,\md t \right |^2.
\end{equation}
Here $\ddot{\bv{d}}(t)$ is the dipole acceleration, which is given by
\begin{equation}
\ddot{\bv{d}}(t) = - \int n(\bv{r},t) \bv{\nabla} H_{\mbox{\scriptsize ks}} \md\bv{r}.
\end{equation}
Note that all the figures in this work present the spectral density along the laser polarization 
direction; $S_k(\omega)$ is negligible along the other directions. 

Additional information can be gained 
by calculating the response for each Kohn-Sham orbital~\cite{chu:2016}. In that case we calculate the 
dipole acceleration from each Kohn-Sham state as
\begin{align}\label{eq:hhg_states}
\ddot{\bv{d}}_j(t) &= - \int n_j(\bv{r},t) \bv{\nabla} H_{\mbox{\scriptsize ks}} \md\bv{r} \nonumber \\ 
&= - 2\int \left|\psi_j(\bv{r}, t)\right|^2 \bv{\nabla} H_{\mbox{\scriptsize ks}} \md\bv{r}\hspace*{0.5cm}j =
1,\dots, N.
\end{align}
While this neglects interferences between different orbitals in the overall harmonic signal, it does give 
an indication of the the contribution of that state.

To gain an insight into the excitation processes occurring during (and following) the VUV pulse, 
we have calculated the overlap of the $N$ time-dependent Kohn-Sham orbitals $\psi_j(\bv{r}, t)$ 
with the lowest-energy occupied and unoccupied field-free Kohn-Sham orbitals, namely
\begin{align}\label{eq:orb_overlap}
 \eta_{k}(t) &= 2 \sum_{j=1}^{N} \left | \int \phi_k^*(\bv{r}) \psi_j(\bv{r}, t) \md\bv{r} \right |^{2}\hspace*{0.2cm}
k = 1,\dots, M,
\end{align}
where $\phi_k(\bv{r})$ are the $M$ ($M > N$) lowest-energy field-free eigenstates of the Kohn-Sham Hamiltonian.

\begin{figure}[t]
\centerline{\includegraphics[width=0.47\textwidth]{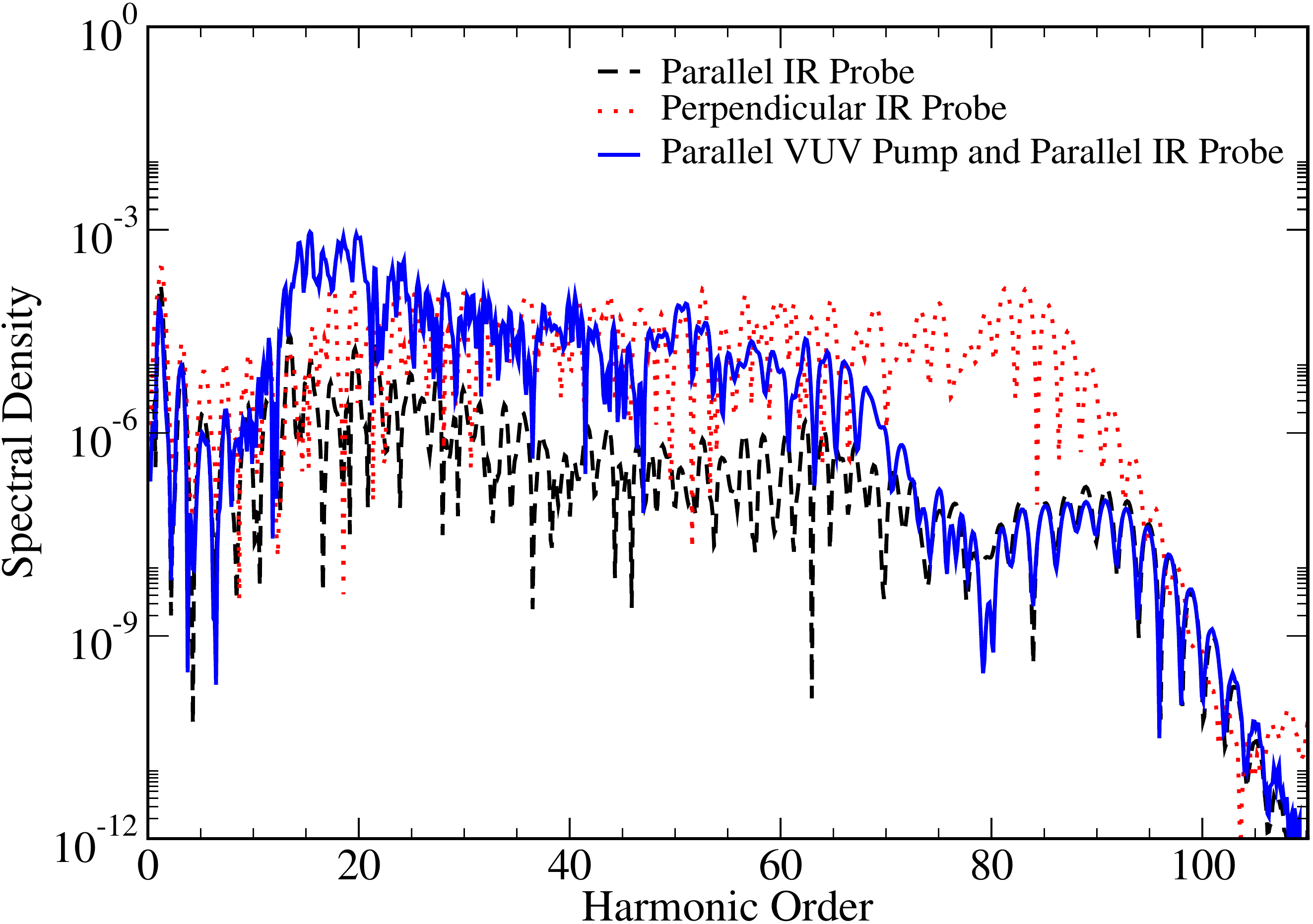}}
\caption{HHG in acetylene after interaction with a 5-cycle, linearly-polarized IR laser pulse having a 
wavelength of $\lambda=$ 1450~nm and a peak intensity of $I =$ \intensity{1.0}{14}. Two orientations of 
the molecule with the field were considered, namely the parallel and perpendicular orientations. Additionally 
we plot the harmonic spectra for the parallel orientation in which the IR pulse has been immediately preceded 
by an 8-cycle, linearly-polarized VUV laser pulse having a wavelength of $\lambda=$ 102~nm and a peak intensity 
of $I =$ \intensity{1.0}{12}.}
\label{fig:figure1}
\end{figure}
\subsection{Static Properties}
Using the pseudopotentials and exchange-correlation functional detailed above, the $X^1 \Sigma_g^+$ ground 
state of the molecule was calculated. This gave a C-C bond length of 2.207~a$_0$ while the C-H bond length 
was 2.045~a$_0$. These agree well with the experimental values of 2.273~a$_0$ and 2.003~a$_0$ 
respectively~\cite{yasuike:2000}. For this equilibrium geometry, both occupied and unoccupied Kohn-Sham 
orbitals were calculated. The energies of these orbitals are shown in Fig.\,\ref{fig:ks_energies}. By 
Koopman's theorem, the magnitude of the HOMO energy can be considered as a good approximation to the vertical 
ionization potential to the $X^2 \Pi_u$ cationic state. Similarly, the magnitudes of the HOMO-1 and HOMO-2 
energies can be thought of as approximations to the energies required to ionize to the $A ^2\Sigma_g^+$ and 
$B ^2\Sigma^+_u$ cationic states respectively. For these three cationic states, the experimental ionization 
potentials (in Hartrees) are 0.4191~Ha, 0.6140~Ha and 0.6912~Ha respectively~\cite{duffy:1992}. Our calculated 
values for these states are 0.4149~Ha, 0.5548~Ha and 0.6117~Ha respectively. 

\begin{figure}[t]
\centerline{\includegraphics[width=0.47\textwidth]{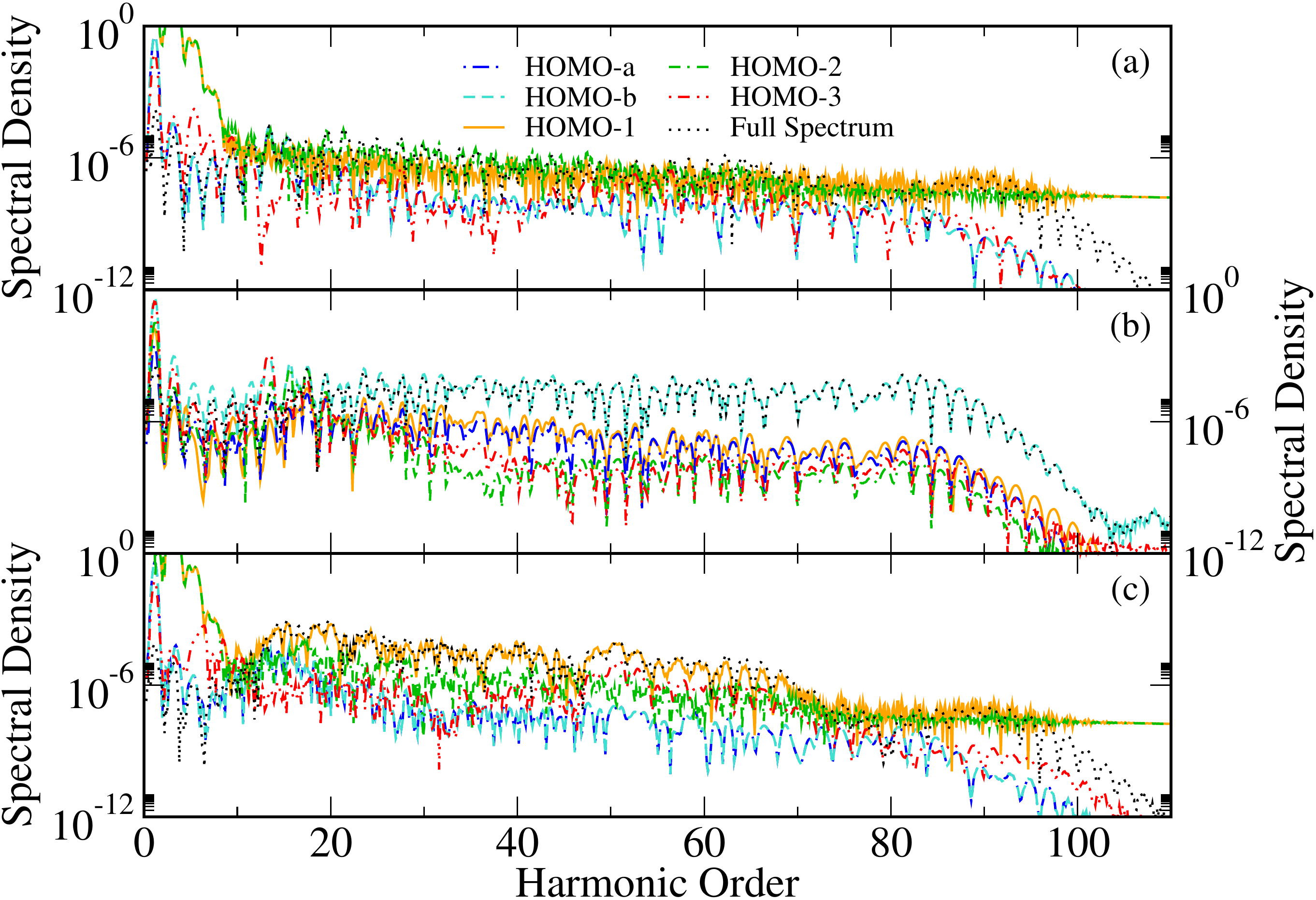}}
\caption{Contributions of individual Kohn-Sham orbitals to the three harmonic spectra shown in 
Fig.~\ref{fig:figure1}, calculated using Eq.\,(\ref{eq:hhg_states}), along with the full harmonic signals for 
reference. Panels (a) and (b) correspond to the parallel and perpendicular orientations respectively with only 
the IR probe present, while panel (c) shows the spectra for the parallel-pump parallel-probe case described 
in the text. In these plots, HOMO-a and HOMO-b refer to the two forms of the HOMO.}
\label{fig:hhga_states}
\end{figure}

\subsection{HHG: IR Pulse}
Consider the interaction of acetylene with a 5-cycle (24.2~fs) laser pulse having a wavelength of 
$\lambda=$ 1450~nm (photon energy $=$ 0.0314 ~Ha) and a peak intensity of $I = $ \intensity{1.0}{14}. 
Figure \ref{fig:figure1} presents harmonic spectra for acetylene aligned both parallel and perpendicular 
to the laser polarization direction. We see that the intensity of the plateau harmonics is several orders 
of magnitude larger when the molecule is aligned perpendicular to the laser polarization direction: this 
well-known result is based on the symmetry of the HOMO. However, two additional features are present in the 
plots. Firstly, for the parallel orientation, we see evidence of a double plateau with an inner plateau 
cut-off near harmonic 65. Secondly, the position of the outer cut-off is different for each orientation.
For the perpendicular orientation, the position of the cut-off is consistent with ionization from the HOMO, 
whereas for the parallel orientation the cut-off is consistent with ionization from the HOMO-1. We note that 
the parallel cut-off is also consistent with ionization from the HOMO-2. We have calculated the HHG spectra 
arising from each occupied Kohn-Sham orbital~\cite{chu:2016}: these are shown in Fig.\,\ref{fig:hhga_states}(a) 
and Fig.\,\ref{fig:hhga_states}(b) for the parallel and perpendicular orientations respectively. These
results suggest that in the perpendicular orientation HHG arises from the HOMO while in the parallel 
case HHG arises predominantly from the HOMO-1, especially in the cut-off region. Extensions to the cut-off 
due to ionization from different orbitals has already been observed in other molecules such as 
N$_2$~\cite{mcfarland:2008}. 

Previous studies of photoabsorption and photoionization in acetylene have considered the role of highly-excited 
states~\cite{machado:1982,han:1989,mitsuke:1994,yasuike:2000,wells:1999,fronzoni:2004}. In particular, features 
in the photoionization spectrum around 0.4963~Ha in the photon energy are generally associated with the 
formation and subsequent autoionization of the $3\sigma_g\rightarrow 3\sigma_u$ excited 
state~\cite{yasuike:2000,wells:1999}. Such a transition is associated with an excitation from the HOMO-1 to the 
LUMO+1. In that case, for HHG in the parallel orientation we could envisage a situation in which an electron
is excited to the LUMO+1, ionized from this state and then recombines back to it. In our simulations, the LUMO+1 
energy is $-$0.1103~Ha and so HHG from the LUMO+1 would give a cut-off at harmonic 77. This does not agree 
with the observed cut-off for the inner plateau around harmonic 65. We will return to this point later. 
\begin{figure}[t]
\centerline{\includegraphics[width=0.47\textwidth]{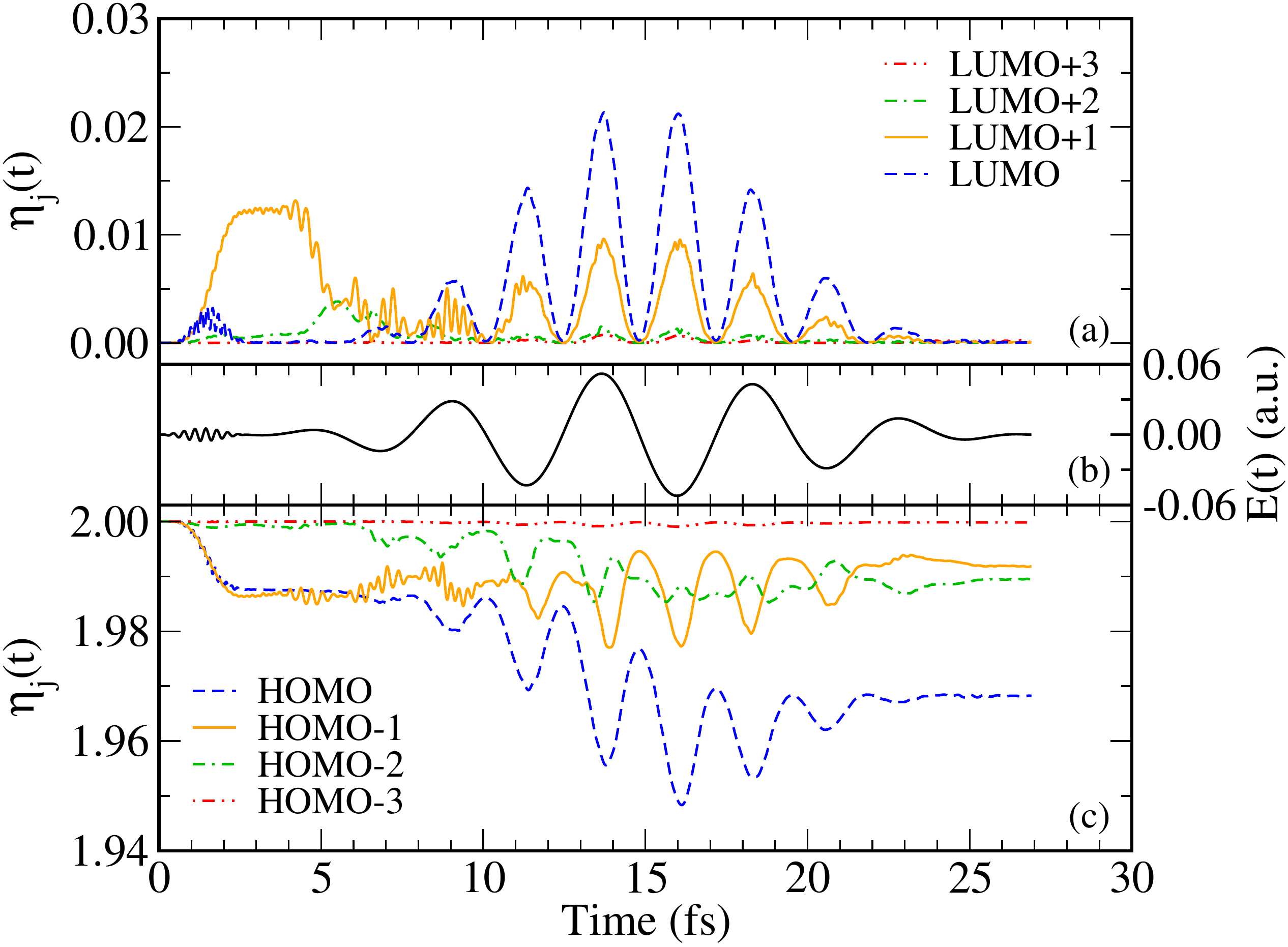}}
\caption{Electronic population in each of the 10 lowest initial Kohn-Sham orbitals $\phi_k(\bv{r})$, 
calculated using Eq.~(\ref{eq:orb_overlap}) during interaction with an 8-cycle, 102\,nm VUV pump pulse with 
peak intensity \intensity{1.0}{12} followed by a 5-cycle, 1450\,nm IR probe with peak intensity 
\intensity{1.0}{14}. Panels (a) and (c) show these populations for the lowest unoccupied and highest occupied 
orbitals respectively. The $z$-component of the electric field vector $\bv{E}(t)$ is provided for reference 
in panel (b). For brevity, only one form of each of the HOMO and LUMO are shown.}
\label{fig:overlaps_vuv_ir}
\end{figure}

\subsection{HHG: VUV Pulse + IR Pulse}
In order to investigate the role of the $3\sigma_g\rightarrow 3\sigma_u$ excitation, we can excite the molecule 
using an 8-cycle (2.67~fs) linearly polarized VUV laser pulse having a wavelength of $\lambda=$ 102~nm (photon 
energy $=0.4467$ Ha) and a peak intensity of $I = $ \intensity{1.0}{12}. For this pulse, the bandwidth is 
sufficient to also excite the $2\sigma_u\rightarrow 4\sigma_g$ transition. Immediately after the pump pulse, the 
molecule interacts with the IR laser pulse. The polarization direction of both pulses is along the molecular axis.
The resulting harmonic spectra is shown in Fig.~\ref{fig:figure1}. Two areas of significant harmonic enhancement 
are observed. Firstly, a window of enhanced harmonics (harmonics 11--21) is present. Secondly, the inner secondary 
plateau that was observed for the IR-only pulse is greatly enhanced. Indeed the intensity of the harmonics in the 
inner plateau are now comparable to those using the IR probe aligned perpendicular to the molecular axis. This 
enhancement could arise from either the $3\sigma_g\rightarrow 3\sigma_u$ or $2\sigma_u\rightarrow 4\sigma_g$ 
excitations. If we calculate the HHG spectra arising from each occupied Kohn-Sham orbital 
(Fig.~\ref{fig:hhga_states}(c)) we see that this plateau arises from the HOMO-1. Furthermore, consider
Fig.~\ref{fig:overlaps_vuv_ir} which shows the calculated overlaps of the time-evolving Kohn-Sham orbitals with the
initial orbitals. Of particular interest are the overlaps with the LUMO+1 and LUMO+2. We clearly see
during the interaction that it is the LUMO+1 that is excited. Hence we attribute the enhancement in the HHG 
spectra to the $3\sigma_g\rightarrow 3\sigma_u$ excitation.

\begin{figure}
\centerline{\includegraphics[width=0.5\textwidth]{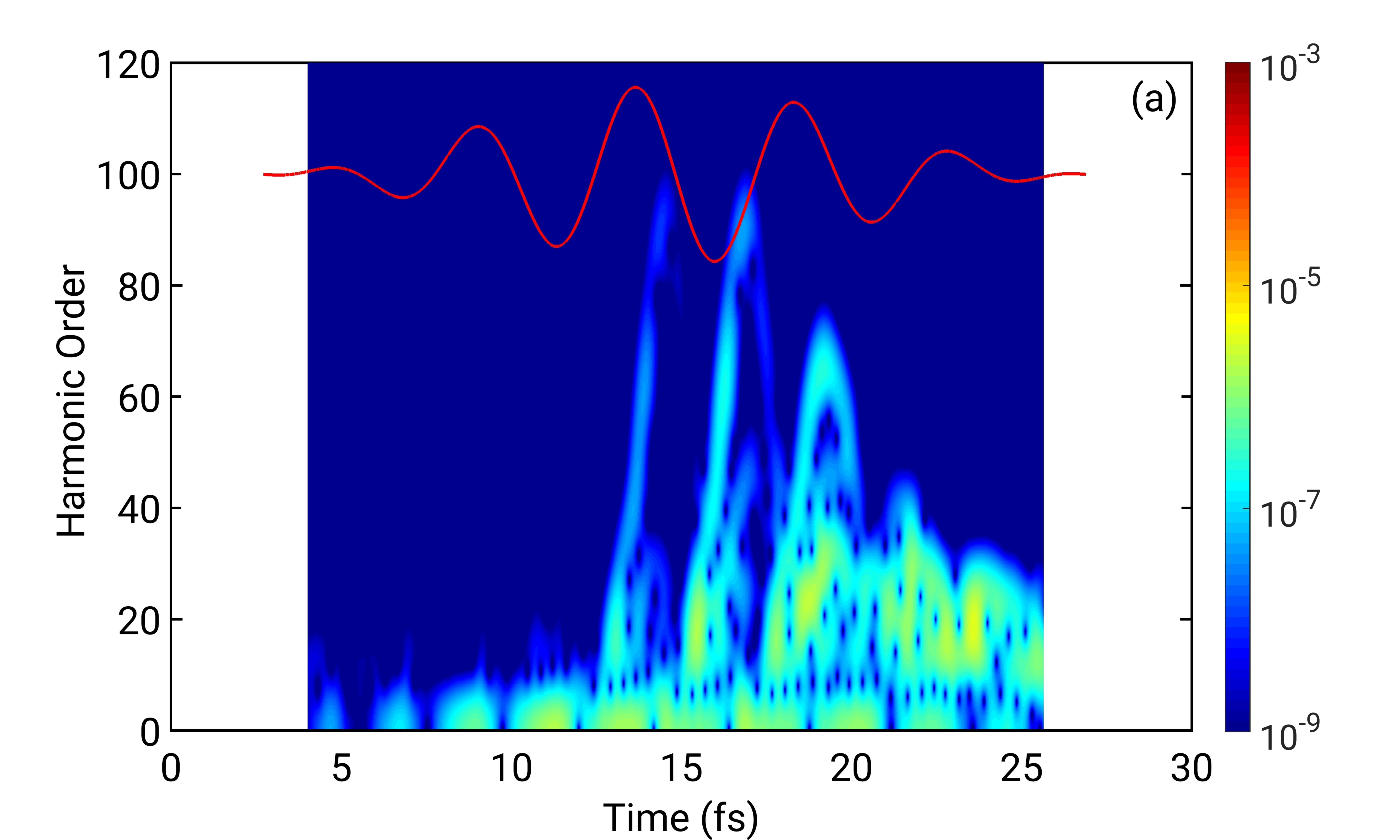}}
\centerline{\includegraphics[width=0.5\textwidth]{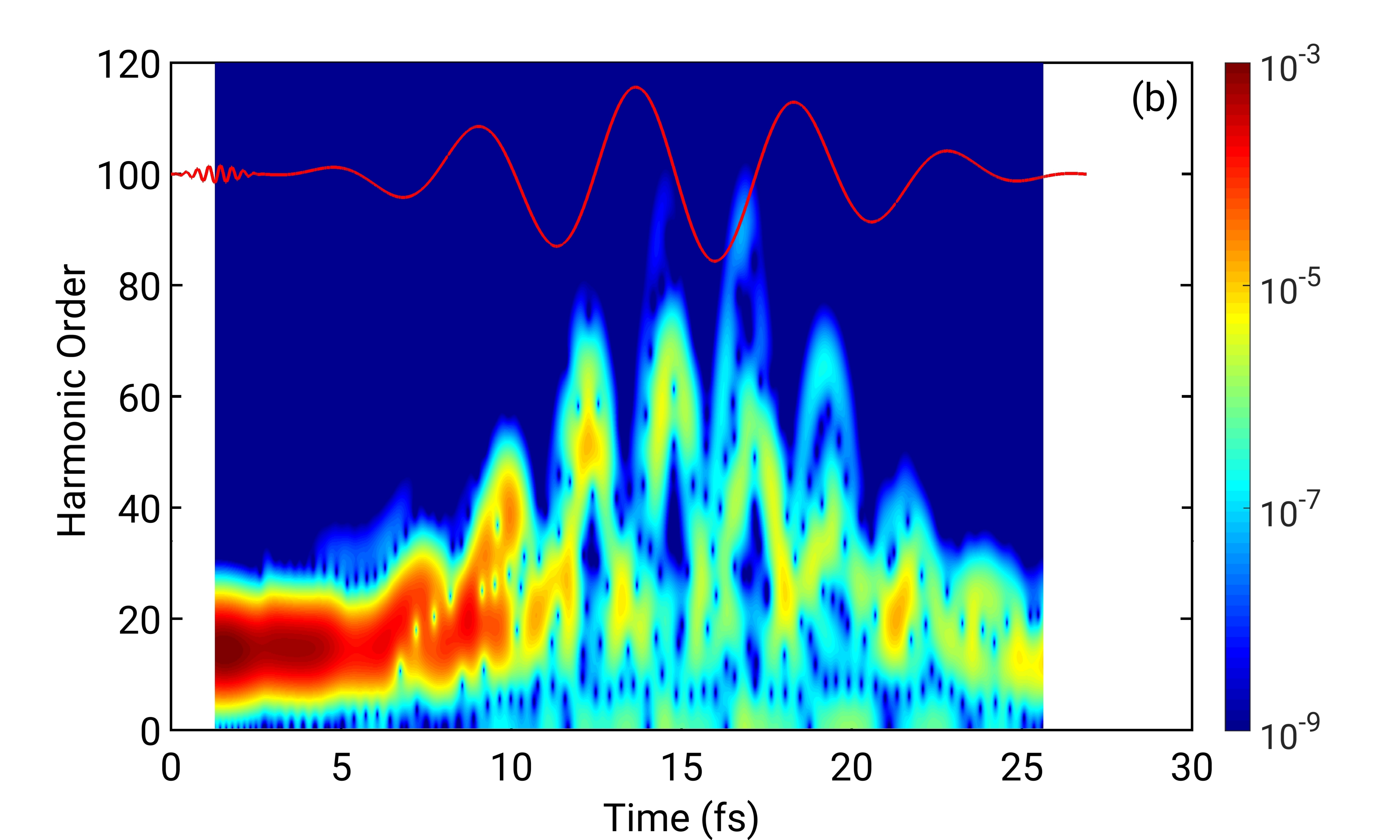}}
\caption{Time frequency analysis of the two HHG spectra obtained in the parallel-orientation in 
Fig.~\protect\ref{fig:figure1}. Plot (a) presents the results obtained using the IR pulse only while plot (b) 
presents those obtained using a VUV pump pulse immediately preceding the IR pulse. The red lines at the top of each plot denote the electric 
field.}
\label{fig:figure2}
\end{figure}

Returning now to the pump-probe spectrum shown in Fig.\,\ref{fig:figure1}, we can identify two mechanisms that 
give rise to the observed harmonic enhancements. After excitation of the molecule by the VUV pump pulse, relaxation 
can now occur back to the ground state resulting in the emission of a high energy photon. For a VUV photon energy of 
0.4467~Ha, this would correspond to harmonic 15 of the IR  pulse. The enhancement for harmonics 11--21 is thus due 
to bound-bound transitions as the molecule relaxes back to the ground state. For the inner plateau, the enhancement 
appears to originate due to ionization from and subsequent recombination back to the excited state. This explanation 
is backed up by carrying out a time-frequency analysis of the harmonic response. Fig.~\ref{fig:figure2}(a) presents 
the time-frequency spectrum for the IR-only pulse in the parallel orientation while Fig.~\ref{fig:figure2}(b) 
presents the results for the VUV+IR pulses. For the IR-only results we see that the plateau harmonics are not 
emitted until the pulse has ramped on fully. This is consistent with the normal three-step model. During the initial 
ramp-on, excitation also occurs which can lead to subsequent HHG from the excited state. For the VUV+IR pulses the 
situation is dramatically different. In this case we see the low-order harmonic enhancement occurring during the 
interaction with the VUV pump right up until the first few cycles of the IR pulse. At this point these bound-bound 
transitions cease as the excited state population ionizes before recombining back to the excited state. These 
bound-continuum transitions occur much earlier during the IR pulse as the population is already excited. Thus we see 
the formation of the plateau well before the pulse has ramped up to maximum intensity. As the IR pulse ramps off, 
HHG switches off and population in the excited state now de-excites to the ground state resulting in the generation 
of low-order harmonics again.

\begin{figure}[t]
\centerline{\includegraphics[width=0.47\textwidth]{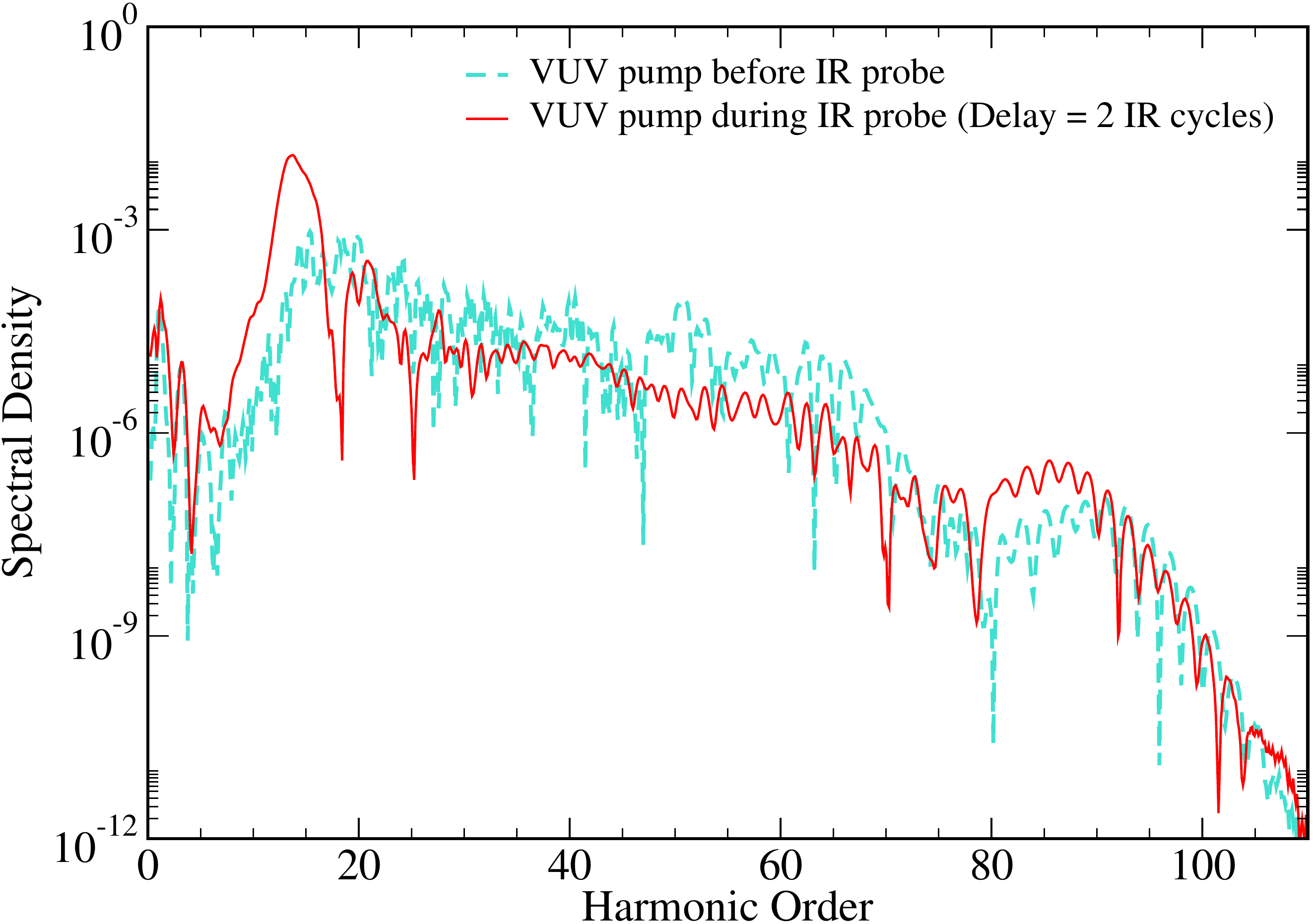}}
\caption{Harmonic spectra from acetylene with different setups of pump and probe pulses.
The VUV pump pulse is an 8-cycle pulse with wavelength $\lambda=$ 102~nm and a peak intensity of 
$I =$ \intensity{1.0}{12}, while the IR probe pulse is a 5-cycle pulse with wavelength $\lambda=$ 1450~nm and 
a peak intensity of $I =$ \intensity{1.0}{14}. All of the pulses are linearly polarized along the molecular axis.}
\label{fig:pump_probe_delay}
\end{figure}

We have also investigated the effect of changing the delay between the VUV pump and IR probe pulses.
The harmonic spectrum resulting from one such calculation is shown in Fig.\,\ref{fig:pump_probe_delay}.
In this case the VUV pulse is applied during the IR pulse, beginning two cycles into the 2 IR pulse (in the following, 
we call this the overlapping pulses setup). We observe less enhancement of the plateau harmonics compared with the case 
where the VUV pulse precedes the IR pulse (we call this the sequential pulses setup). This is because a VUV pulse 
applied to the ground state molecule before the IR pulse is much more effective at exciting the 
$3\sigma_g\rightarrow 3\sigma_u$ transition than one applied during the IR pulse, when the system has already been 
excited. The window of harmonics due to bound-bound transitions also differs for the two setups as shown in
Fig.\,\ref{fig:pump_probe_delay}. In the sequential case, the VUV pulse is applied when the molecule is in its ground 
state. During the interaction with this pump pulse, the LUMO+1 is populated, with the subsequent transitions back to the 
ground state producing relatively well defined harmonics. For the overlapping pulses, the situation is different. Now 
the VUV pulse is being applied when the system is already far from equilibrium, and where some population may have 
been excited to the LUMO+1 by the IR pulse. Applying the VUV pump in this case can excite more transitions to the 
LUMO+1, as before, but can also stimulate the emission of the low-order (bound-bound) harmonics. The broadening of the 
peak reflects the distorted nature of the molecule at the time when the VUV pulse is applied.

Rather than overlapping the two pulses, we could instead increase the delay time between them. As might be expected,
the spectra produced in these cases (not shown here) are similar to the spectrum produced for sequential pump and 
probe pulses, but with the harmonics due to bound-bound transitions increasing in both sharpness and intensity. 
At very long pump-probe delays, the enhancement of the plateau decreases.

\begin{figure}[t]
\centerline{\includegraphics[width=0.47\textwidth]{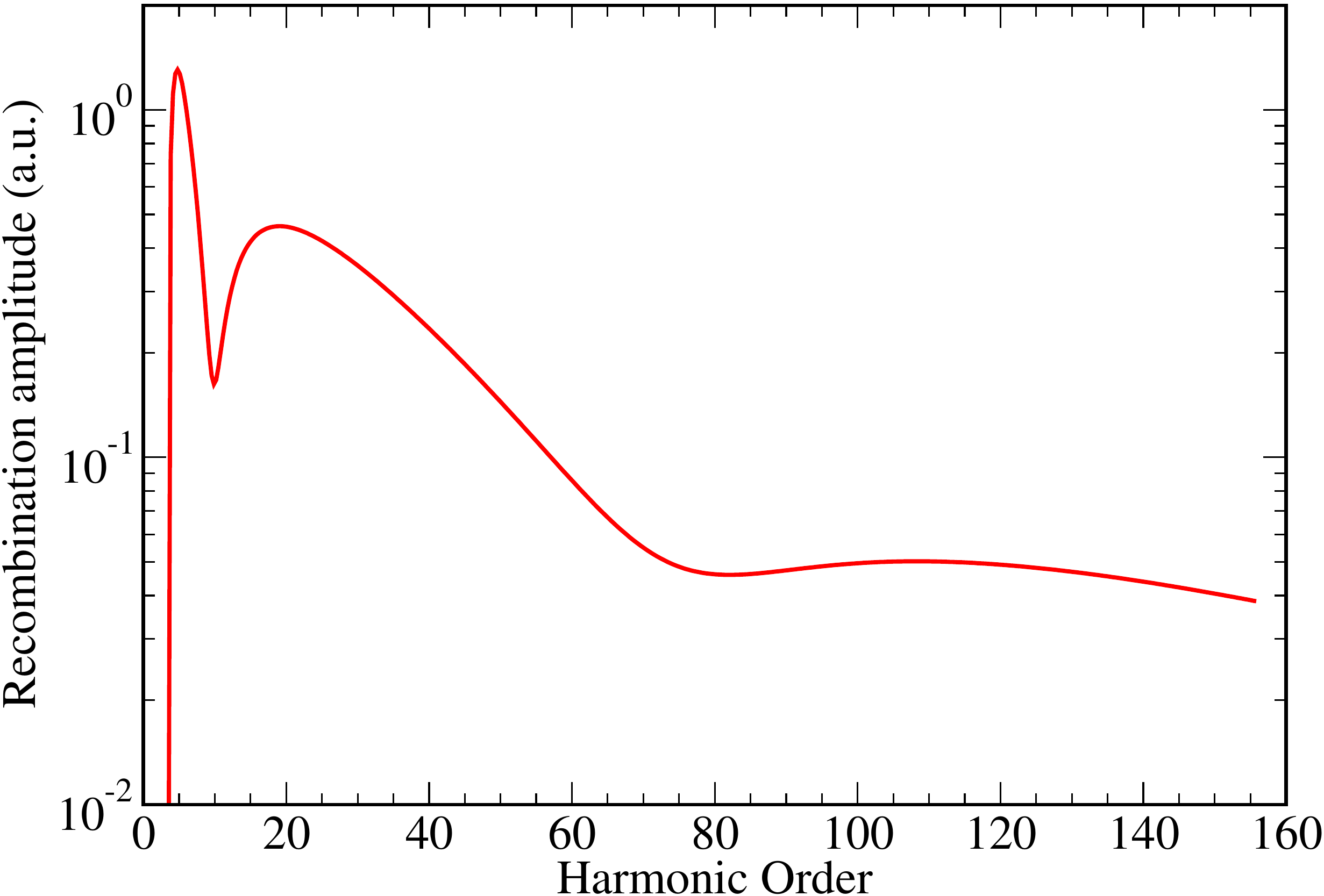}}
\caption{Dipole recombination matrix element for a continuum electron recombining back to  the LUMO+1 autoionizing 
state. The velocity form of the recombination element is calculated as described in Eq.~\ref{eq:eq1}.
The energy of the free electron and the ionization potential of the Kohn-Sham state are related to a given harmonic 
order, $n$, using the classical formula $n\omega_L = k^2/2 + I_p$, where $\omega_L$ is the frequency of our IR field 
and $I_p$ is the ionization potential of the LUMO+1.}
\label{fig:figure3}
\end{figure}

\begin{figure}[t]
\centerline{\includegraphics[width=0.47\textwidth]{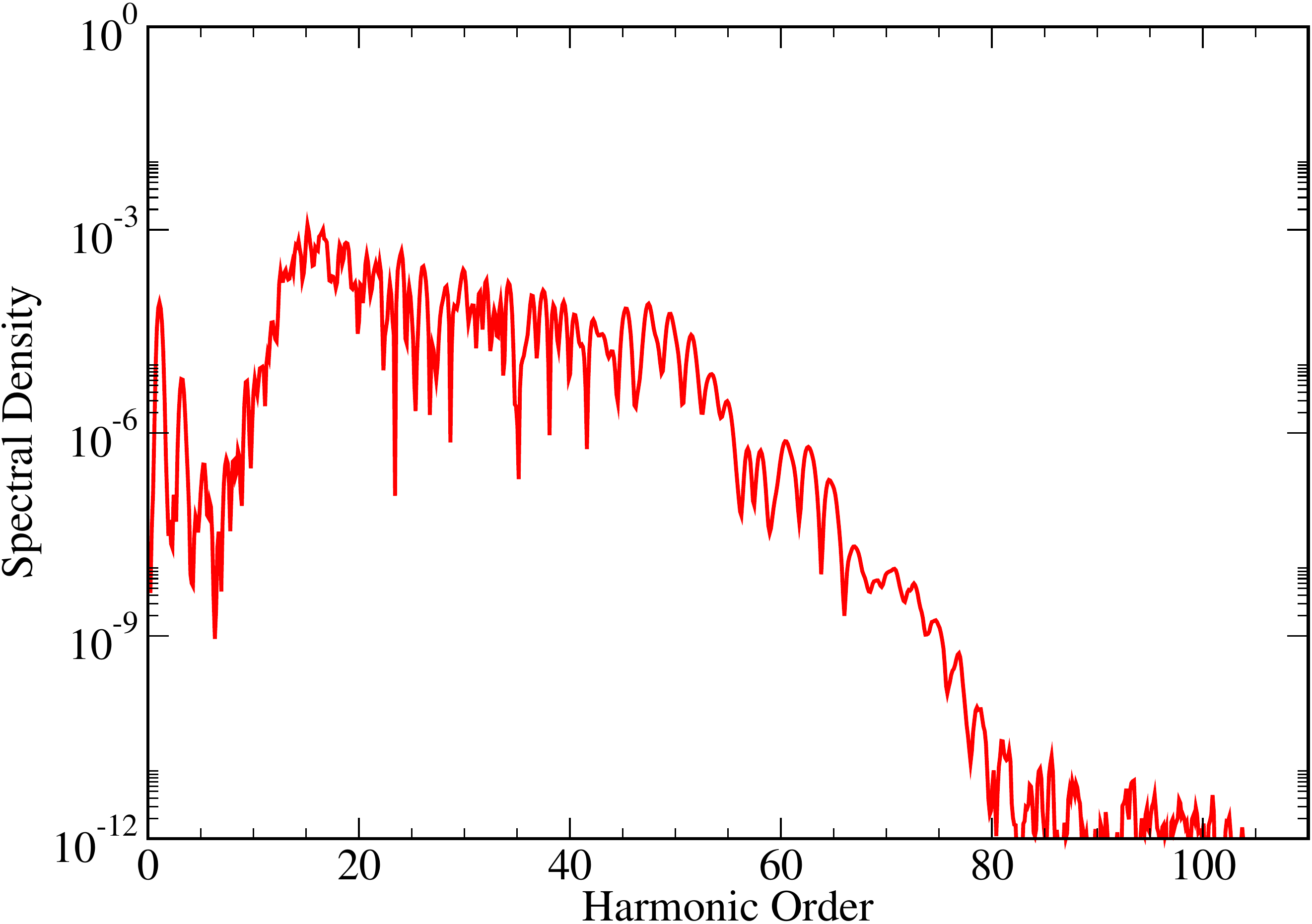}}
\caption{HHG in acetylene after interaction with a VUV pump pulse and IR probe pulse. The pump pulse is an 8-cycle 
linearly-polarized VUV laser pulse having a wavelength of $\lambda=$ 102~nm and a peak intensity of 
$I =$ \intensity{1.0}{12}. The probe pulse is a 5-cycle, linearly-polarized IR laser pulse having a wavelength of 
$\lambda=$ 1450~nm and a peak intensity of $I =$ \intensity{6.5}{13}.}
\label{fig:figure4}
\end{figure}

\subsection{HHG: Position of inner plateau cut-off}
Finally, we return to the question of the position of the cut-off for the excited-state plateau: in essence there 
appears to be a suppression of the cut-off harmonics. Minima in harmonic spectra are a well studied feature in HHG.
A number of potential mechanisms could give rise to minima including 
multi-channel~\cite{smirnova:2009,woerner:2010,haessler:2010,miao:2014} or structural 
interference~\cite{lein:2002a,lein:2002b,itatani:2004,kanai:2005,vozzi:2005,le:2006,torres:2007,kato:2011,etches:2011} 
effects. Structural interferences can be studied by calculating recombination matrix 
elements~\cite{ciappina:2007,etches:2011,zu:2011}. The recombination matrix element of an ionized electron to a 
bound state of the molecule can be written as
\begin{equation}
\bv{d}_{\mbox{\scriptsize rec}}(\bv{k}) = \left<\chi(\bv{k},\bv{r})\bigr|-i\bv{\nabla}\bigr|\Psi(\bv{r})\right>,
\label{eq:eq1}
\end{equation}
in the velocity form where $\chi(\bv{k},\bv{r})$ is a plane wave describing the free electron having momentum 
$\bv{k}$ while $\Psi(\bv{r})$ is the wavefunction of the bound state. We can calculate the absolute value of this 
amplitude in a given direction $\bv{e}$ due to a recombination back to any of the field-free Kohn-Sham states of 
the molecule: in this work we take $\bv{e}$ to be along the laser polarization direction. Integrating this over 
all angular variables for $\bv{k}$ and relating the energy of the free electron and the ionization potential of 
the Kohn-Sham state to a given harmonic order, $n$, using the classical formula gives $n\omega_L = k^2/2 + I_p$, 
where $\omega_L$ is the frequency of our IR field and $I_p$ is the ionization potential of the state, 
$\Psi(\bv{r})$. We plot this recombination amplitude for the LUMO+1 in Fig.~\ref{fig:figure3}. Immediately 
we see that there is a minima in the recombination amplitude at harmonic 81, close to the predicted cut-off for 
recombination to the LUMO+1 for the chosen laser parameters. Suppose we consider the same interaction, but 
now with the intensity of the IR pulse lowered to $I =$ \intensity{6.5}{13}. In that case the classical cut-off 
should be at harmonic 53, away from the minimum in the the recombination amplitude to the LUMO+1. The harmonic 
spectra for this case is shown in Fig.~\ref{fig:figure4} and we see that the observed cut-off agrees with the 
predicted value. Thus our results support the idea that the main features of this secondary plateau arise due to 
ionization from and subsequent recombination to the LUMO+1.

\section{Conclusions}
\label{sec:conclusions} 
In conclusion, we have studied HHG in aligned acetylene molecules using mid-IR laser pulses and shown that an 
autoionizing state associated with the $3\sigma_g\rightarrow 3\sigma_u$ excitation plays a crucial role whenever 
the pulse is aligned along the molecular axis. By exciting the molecule with a VUV pulse tuned to this excitation, 
the harmonic signal in the plateau is dramatically enhanced. This enhancement is associated with ionization from 
and recombination back to the excited state. Since resonances and autoionizing states are ubiquitous in many molecules 
this opens up the possibility of controlling reactions using suitable attosecond pulses~\cite{krauz:2009,lepine:2014}. 
Additionally, since the HHG enhancement using the VUV+IR pulses is comparable to the spectral density observed using 
an IR pulse aligned perpendicular to the molecule and since the pump pulse does not appear to have an effect on the 
response in the perpendicular alignment, the use of the VUV pump should lead to a boost in the overall harmonic signal 
when considering unaligned samples of molecules.

\begin{acknowledgments} 
This work used the ARCHER UK National Supercomputing Service (http://www.archer.ac.uk) and has been supported by 
COST Action CM1204 (XLIC). PM acknowledges financial support through a PhD studentship funded by the UK Engineering 
and Physical Sciences Research Council.
\end{acknowledgments} 


\end{document}